\documentstyle[epsfig,12pt]{article}

\begin{document}

\begin{center}

{\bf
\title "The apsidal motion in binary stars.}

\bigskip

\bigskip
\author "B.V.Vasiliev
\bigskip

{Institute in Physical-Technical Problems, 141980, Dubna, Russia}
\bigskip

{vasiliev@dubna.ru}
\end{center}

\bigskip

\begin{abstract}
It is usually accepted to consider an apsidal motion in binary
stars as a direct confirmation that a substance inside stars is
not uniformly distributed. It is shown in this paper that the
apsidal motion in binary systems observation data is in a good
agreement with an existence of uniform plasma cores inside stars
if they consist of hydrogen-deuterium-helium mixture.

\end{abstract}
\bigskip

\section{}

An apsidal motion (an advance of periastron) of close binary stars
is a result of their non-keplerian moving. It originates from a
non-spherical form of stars. The last has been produced by a
rotation of stars or their mutual tidal effects. The traditional
approach to the estimation of a value of this effect needs to
suppose some redistribution of mass inside stars to reach an
accordance between the measured data and calculations. Usually it
is necessary to assume that the density of a substance at the
central region of a star is hundred times greater than a mean
density of a star \cite{peri1}, \cite{peri2}.

On the face of it, the model of constant parameter
\cite{BV-book},\cite{BV}, where density inside a star core is
constant, is in direct contradiction with the data of observation
of an apsidal motion in close binary systems \cite{Hall}. However,
the traditional solution of the task about an apsidal motion is
obtained for electrically non-charged matter. It is obvious that
for electrically charged matter this solution must be
reconsidered.

Let us estimate a contribution of inherent rotation of stars in an
apsidal motion. According to \cite{peri1}-\cite{peri2}, the ratio
of $\omega$ the angular velocity of an apsidal motion and $\Omega$
the angular velocity of rotation of a star about its axis is

\begin{equation}
\frac{\omega}{\Omega}=\frac{3}{2}\frac{(I_A-I_C)}{Ma^2},
\end{equation}

where $I_A$ and $I_C$ are momenta of inertia relative to principal
axes of the ellipsoid:

\begin{equation}
I_A-I_C=\frac{M}{5}(a^2-c^2),
\end{equation}

where $a$ and $c$ are polar and equatorial radii of stars.

Thus

\begin{equation}
\frac{\omega}{\Omega}=\frac{3}{10}\frac{(a^2-c^2)}{a^2}\label{oo}.
\end{equation}

It is correct when an angular velocity on the ellipse coincides
with $\Omega$ an angular velocity of rotation of a star about its
axis .

In the absence of rotation, the equilibrium equation  of plasma
inside a star is \cite{BV}:

\begin{equation}
\gamma g_G+\rho_G E_G=0
\end{equation}

where $g_G$, $\rho_G$ and $E_G$ are acceleration of gravitation,
gravity-induced density of charge and intensity of gravity-induced
electric field. ($div~g_G=4\pi~ G~ \gamma$, $div~E_G=4\pi \rho_G$
and $\rho_G=\sqrt{G}\gamma$).

\section{}

One can suppose that at rotation, an additional electric charge
with density $\rho_\Omega$ and electric field $E_\Omega$ can arise
under action of a rotational acceleration  $g_\Omega$, and the
equilibrium equation takes the form:

\begin{equation}
(\gamma_G+\gamma_\Omega)(g_G+g_\Omega)=(\rho_G+\rho_\Omega)(E_G+E_\Omega),
\end{equation}

where

\begin{equation}
div~(E_G+E_\Omega)=4\pi(\rho_G+\rho_\Omega)
\end{equation}

or

\begin{equation}
div~E_\Omega=4\pi\rho_\Omega.
\end{equation}

We can seek a solution for electric potential as an expansion on
spherical functions. Assuming that an eccentricity is small, we
can limit by a second term of the expansion:

\begin{equation}
\varphi=C_\Omega~r^2(3cos^2\theta-1)
\end{equation}

or in Cartesian coordinates

\begin{equation}
\varphi=C_\Omega(3z^2-x^2-y^2-z^2),
\end{equation}

where $C_\Omega$ is a constant.

 Thus

\begin{equation}
E_x=2~C_\Omega~x,~ E_y=2~C_\Omega~y,~ E_z=-4~C_\Omega~z
\end{equation}

and

\begin{equation}
div~E_\Omega=0
\end{equation}

we obtain important equations:

\begin{equation}
\rho_\Omega=0;
\end{equation}

\begin{equation}
\gamma g_\Omega=\rho E_\Omega.
\end{equation}

Since a centrifugal force must be counterbalanced by the electric
force

\begin{equation}
\gamma~\Omega^2~x=\rho~2C_\Omega~x
\end{equation}

the constant is

\begin{equation}
C_\Omega=\frac{\gamma~\Omega^2}{2\rho}=\frac{\Omega^2}{2\sqrt{G}}
\end{equation}

The potential of a positively uniformly charged ball is

\begin{equation}
\varphi(r)=\frac{Q}{R}\biggl(\frac{3}{2}-\frac{r^2}{2R^2}\biggr)
\end{equation}

A negative surface charge induces inside the sphere an electric
potential

\begin{equation}
\varphi(R)=-\frac{Q}{R},
\end{equation}

where according to \cite{BV} $Q=\sqrt{G}M$, $M$ is the mass of a
star.

Thus, the total potential inside the star under consideration is

\begin{equation}
\varphi_\Sigma=\frac{\sqrt{G}M}{2R}\biggl(1-\frac{r^2}{R^2}\biggr)+\frac{\gamma
\Omega^2}{2\rho_G}r^2(3cos^2\theta-1)
\end{equation}

The electric potential must be equal to zero on the surface of a
star, since at $\varphi_\Sigma=0$ and at $r=R$ we obtain

\begin{equation}
R^2-R_0^2=\frac{\gamma\Omega^2}{\frac{4\pi}{3}\rho^2},
\end{equation}

where $R_0$ is the radius of a star without rotation. This
equation describes the form of a rotating star. From this equation

\begin{equation}
a^2-c^2=\frac{\Omega^2}{G\gamma}\frac{9}{4\pi}.
\end{equation}

and Eq.({\ref{oo}}) gives

\begin{equation}
\frac{\omega}{\Omega}=\frac{27}{40\pi}\frac{\Omega^2}{G\gamma}.
\end{equation}

When both stars of a close pair induce an apsidal motion, this
equation transforms to

\begin{equation}
\frac{\omega}{\Omega}=\frac{27}{40\pi}\frac{\Omega^2}{G}\biggl(\frac{1}{\gamma_1}+\frac{1}{\gamma_2}\biggr),\label{1}
\end{equation}

where $\gamma_1$ and $\gamma_2$ are densities of stars.

\section{}

A density of non-degenerate non-relativistic plasma can be
determined in the following way. At high density and temperature
an electron gas is almost ideal \cite{LL8} and the ideal gas law
will serve as its equation of state in first approximation. It
needs to take to account two corrections to describe electron gas
properties more exactly  - a correction for identity of electrons
and a correction for a presence of positively charged nuclei. The
first of them is positive, it increases the incompressibility of
electron gas. The second correction decreases its
incompressibility and it is negative. These correction are known
\cite{LL8},and that is why with allowance for both corrections the
free energy of electron gas takes the form

\begin{equation}
F=F_{ideal}+N\frac{\pi^{3/2}e^3 a_0^{3/2}}{4(kT)^{1/2}}n -
N\frac{2 e^3 Z^3}{3}\frac{\pi^{1/2}}{(kT)^{1/2}}n^{1/2}, \label{}
\end{equation}

where $Z$ is the nucleus charge,$a_0=\frac{\hbar^2}{ me^w2}$ is
the Bohr radius.

At a constant full number of particles in the system and at
constant temperature, the free energy minimum equation is

\begin{equation}
\biggl(\frac{\partial F}{\partial n}\biggr)_{N,T}=0, \label{}
\end{equation}

what allows one to obtain the steady-state value of density of
particles of hot non-relativistic plasma

\begin{equation}
n_0=\frac{16Z^6}{9 \pi^2 a_0^3}\simeq 2\cdot 10^{24} Z^6 ~
cm^{-3}. \label{}
\end{equation}

Thus, the equilibrium density of a star is

\begin{equation}
\gamma=\frac{16Am_pZ^5}{9\pi^2a_0^3}\label{26}
\end{equation}

where $A$ is the mass number, $m_p$ is proton mass.

\section{}

If we introduce $P=\frac{2\pi}{\Omega}$ the period of ellipsoidal
rotation and $U=\frac{2\pi}{\omega}$ the period of the apsidal
motion,  we obtain from Eq.({\ref{1}}) the main equation of our
theory:

\begin{equation}
\frac{P^3}{U}\xi=\biggl(\frac{1}{A_1Z_1^5}
+\frac{1}{A_2Z_2^5}\biggr),\label{2}
\end{equation}

where the constant

\begin{equation}
\xi=\frac{243~\pi^3~a_0^3}{160~G~m_p}~sec^{-2}.
\end{equation}

For hydrogen the value $1/AZ^5$ is equal to 1, for deuterium 0.5,
for helium 1/128. Thus, for two stars both consisting of hydrogen
the right part of Eq.({\ref{2}}) is equal to 2. For  pairs
consisting of hydrogen-deuterium it is 1.5. For D-D pairs and H-He
pairs it equals to 1. For D-He pairs 0.5, and for He-He pairs
1/64.

Thus, for pairs of a star consisting of hydrogen, deuterium and
helium

\begin{equation}
\frac{1}{64}<\frac{P^3}{U}\xi<2.\label{}
\end{equation}

\begin{figure}
\begin{center}
\includegraphics[12cm,2cm][18cm,12cm]{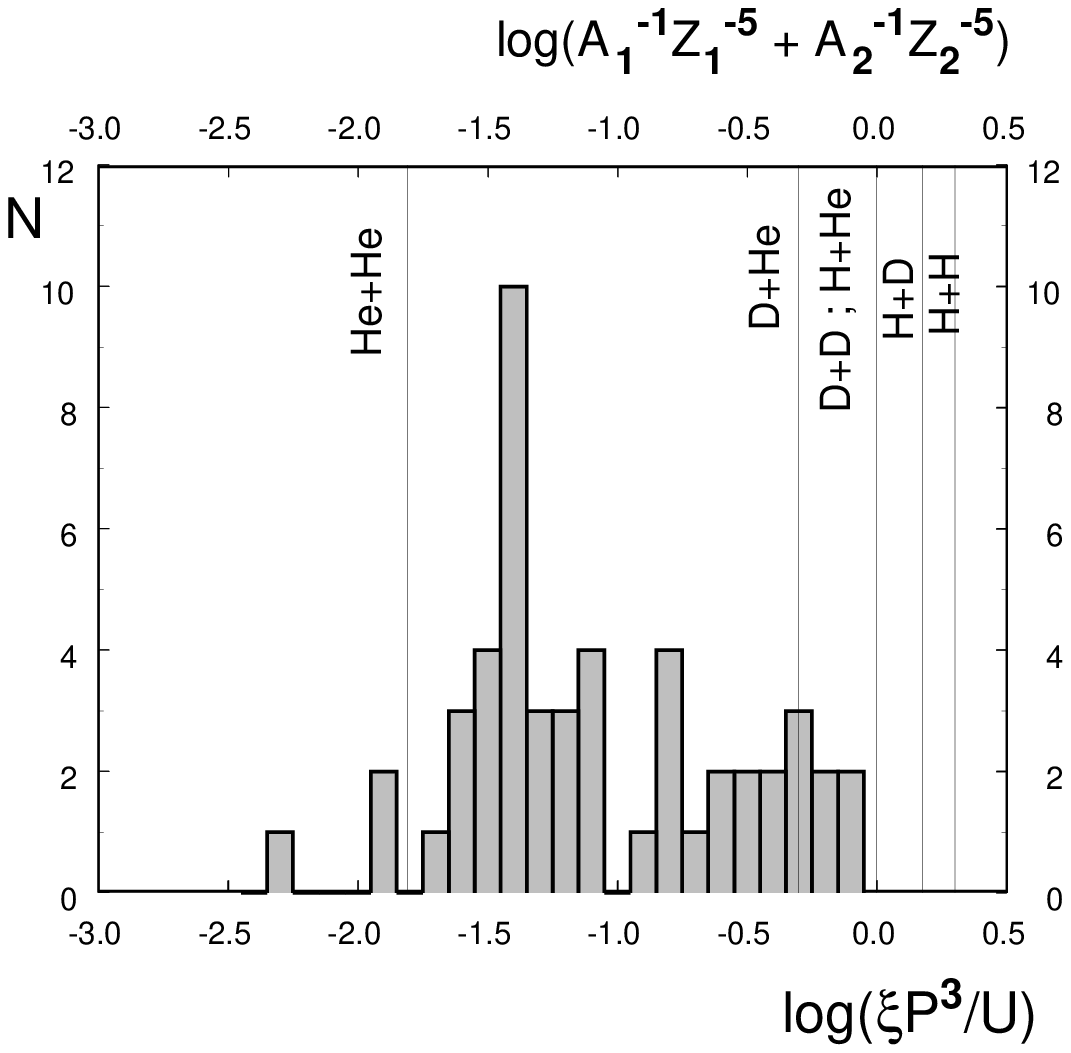}
\caption {The distribution of binary stars on value of
$P^3\xi/U$.} \label{periastr}
\end{center}
\end{figure}

The periods $U$ and $P$ are measured for few tens of stars
\cite{Hall} and we can compare our calculation with the data of
these measurements. The distribution of binary stars by the value
of $P^3\xi/U$ is shown on Fig.{\ref{periastr}} in logarithmic
scale. On upper abscissa the value of $log\big(\frac{1}{A_1
Z_1^5}+\frac{1}{A_2 Z_2^5}\big)$ is shown. The lines mark the
values of parameters for different pairs of binary stars.

It can be seen that all measured data (with the exception of few
pairs only) are limited by factors 1 and 1/64. This argues for
adequate interpretation and satisfactory accuracy of our
estimations of the effect of the apsidal motion.

The interesting result can be obtained if we additionally take in
to account the measured data on mass of rotating pairs.
Accordingly \cite{BV}, the star mass is determined by the $A/Z$
ratio:

\begin{equation}
\frac{A}{Z}=\sqrt{\frac{12 M_\odot}{M_{\star}}}
\end{equation}

and from Eq.{\ref{26}}

\begin{equation}
A~Z^5=\frac{\gamma_{\star} 9\pi^2 a_0^3}{16 m_p}
\end{equation}

\begin{figure}
\begin{center}
\includegraphics[12cm,6cm][18cm,16cm]{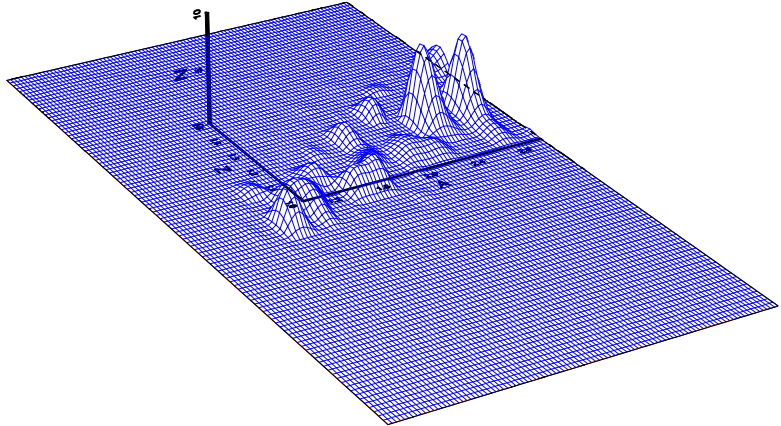}
\caption {The distribution of binary stars on values A and Z.}
\label{dis-az}
\end{center}
\end{figure}

At solution of this equations with using of data observation of
$M_{\star}$ and $\gamma_{\star}$ from the apsidal motion we can
obtain a distribution of stars on $Z$ and $A$ separately
(Fig.{\ref{dis-az}}). As the periods of the apsidal motion of
stars consisting of nuclei heavier than helium are more than
$10^5$ years and rather are immeasurable, this distribution is cut
off on $1<Z<2$ and $1<A<3$. This distribution rather does not give
a new astrophysical information, but it illustrates an efficiency
and a cardinal utility of developed consideration.

\begin{thebibliography}{5}
\bibitem {peri1} Russel H.N., Monthly Notices of the RAS {\bf{88}} (1928) 642
\bibitem {peri2} Chandrasekhar S., Monthly Notices of the RAS {\bf{93}} (1933) 449
\bibitem {BV-book} Vasiliev B.V. - The Physical Approach to Stellar Classification, 2000, Grant Publisher,Moscow.
\bibitem {BV}   Vasiliev B.V. - Nuovo Cimento B, {\bf{116}} (2001) 617-634.
\bibitem {Hall}   Халиуллин Х.Ф. Двойные звезды, Москва, Косминформ, 1997.
\bibitem {LL8}    Landau L.D. and Lifshits E.M. - Statistical Physics,1980, vol.1, 3rd edition,Oxford:Pergamon.



\end{thebibliography}
\end{document}